\documentclass{anstrans}
\title{Algorithms for Achieving Subpixel Resolution in Muon Tomography}
\author{Matthew Mark Romano,$^{1,2}$ JungHyun Bae,$^{2,*}$ Paul Cantonwine$^{2}$}

\institute{
$^{1}$Florida Institute of Technology, 150 W University Blvd, Melbourne, FL 32901\\
$^{2}$Oak Ridge National Laboratory, 1 Bethel Valley Rd, Oak Ridge, TN 37830
}

\usepackage{graphicx}
\usepackage{booktabs}
\usepackage{microtype}
\usepackage{amsmath}
\usepackage{siunitx}
\usepackage{url}
\usepackage{fancyhdr} 
\usepackage{subfigure} 

\fancypagestyle{firstpagefooter}{
    \fancyhf{} 
    \fancyfoot[L]{%
    \parbox{\textwidth}{\footnotesize 
    \vspace{1mm}
    *Corresponding author: baejh@ornl.gov.\\[2mm]
    This manuscript has been authored by UT-Battelle, LLC, under contract DE-AC05-00OR22725 with the US Department of Energy (DOE). The US government retains and the publisher, by accepting the article for publication, acknowledges that the US government retains a nonexclusive, paid-up, irrevocable, worldwide license to publish or reproduce the published form of this manuscript, or allow others to do so, for US government purposes. DOE will provide public access to these results of federally sponsored research in accordance with the DOE Public Access Plan (\url{http://energy.gov/downloads/doe-public-access-plan}).
    }}
}

\begin{document}

\thispagestyle{firstpagefooter}

\section{Introduction}

Muon tomography has emerged as a promising nondestructive imaging technique for monitoring the internal structure of dense and shielded objects by exploiting the natural flux of cosmic ray muons. Unlike conventional radiographic methods, which rely on transmitted photons or neutrons, muon imaging utilizes multiple Coulomb scattering (MCS) to infer material composition and geometry. This use of MCS means that muon tomography offers a unique advantage in nuclear energy and nuclear security applications, particularly for inspecting spent nuclear fuel canisters, verifying the integrity of dry storage casks \cite{doi:10.1098/rsta.2018.0052, Bae2024}, and detecting the illicit trafficking of nuclear materials \cite{8069919, BAE2024110240}.

A critical limitation to the practical application of muon tomography systems is the spatial resolution of the tracking detectors---typically, scintillators or gaseous detectors with discrete readout granularity \cite{MOSES2011S236}. The finite pitch or segmentation of these physical detector elements inherently constrains the positional accuracy with which muon trajectories can be reconstructed. To transcend this hardware-imposed limit, subpixel resolution techniques become necessary. Achieving subpixel resolution---reconstructing muon positions with accuracy finer than the detector's physical segmentation---is thus necessary for advancing the fidelity of muon imaging. However, subpixel positioning in muon detectors poses technical challenges. Muons are minimum ionizing particles that deposit limited energy in each interaction, and scintillation light yields can be sparse and variable. Additionally, light propagation within the scintillators introduces spatial diffusion, complicating the estimation of true interaction position.

The pursuit of subpixel position reconstruction is especially valuable because it will enable significant performance gains using existing detector technologies and will thereby prevent the need for costly and complex hardware modifications. Such algorithmic enhancements are critical for overcoming the fundamental challenge of low cosmic muon flux at sea level, which limits imaging speed. By maximizing the accuracy of each measurement, these methods can make muon tomography a more practical and timely inspection tool.

In this work, we investigated three computational approaches to achieve subpixel resolution in a simulated muon tomography system using a scintillator:

(1) Centroid reconstruction method: a computationally lightweight technique that estimates the muon hit position based on the energy distribution across neighboring scintillator cells.

(2) Maximum likelihood estimation (MLE): a statistically grounded method that seeks the most probable interaction position given the observed energy profile and detector response.

(3) ML approach: a feedforward neural network trained on simulated data to infer muon positions based on engineered features derived from energy deposition patterns.

Using a Geant4-based model of a four-plane scintillator system and 4 GeV muons, we evaluated each method's ability to resolve muon interaction positions within a single detector plane. Our results indicate that all methods achieve subpixel localization; the ML method yields the highest accuracy, reducing mean positional error to just over 10\% of the detector cell size. These findings support the potential of algorithmic advancements to substantially improve muon tomography performance without the need to modify existing detector hardware. Such innovations directly align with national priorities in cost-effective nuclear safeguards, nonproliferation, and radioactive waste monitoring, offering scalable improvements for next-generation imaging systems.

\section{Multiple Coulomb Scattering}

Cosmic ray muon tomography relies on detecting the trajectories of naturally occurring muons to image dense materials. A typical muon tomography apparatus consists of two pairs of position-sensitive particle detectors installed before and after a scanning volume. Each pair of detectors records the 2D position of the muon. These position measurements are combined to reconstruct muon trajectories within the scanning volume. Because muons are charged particles, they undergo consecutive deflection when they interact with matter due to Coulomb interaction with nuclei and electrons. This phenomenon is known as MCS, and it is directly proportional to the atomic number (Z) of the material. The expected scattering angle distribution follows the Gaussian distribution with a zero mean, with the RMS projected angle $\theta_0$ is given by Eq.~(1) \cite{PDG2020}:

\begin{equation}
\theta_0 = \frac{13.6\,\text{MeV}}{\beta cp} \sqrt{\frac{x}{X_0}} \left[1 + 0.038 \ln\left(\frac{x}{X_0}\right)\right]
\end{equation}

\noindent where $\beta$ is the particle velocity in units of $c$, $p$ is the momentum in MeV/c, $x$ is the thickness of the material, and $X_0$ is the radiation length of the material (which scales approximately as $A/Z^2$, where $A$ is the atomic mass and $Z$ is the atomic number).

Although this analytical model aligns well with experimental results, low surface-level cosmic muon flux motivates more accurate reconstruction methods. Because practical detector configurations are often limited by pixel size, this coarse resolution limits imaging capabilities, motivating the development of subpixel reconstruction techniques.

\section{Investigated Method Improvements}

\subsection{Detector Configuration and Simulation}
The muon tomography system is modeled using Geant4 version 11.3.2, employing the FTFP\_BERT physics list with optical physics enabled for scintillation modeling. A typical muon tomography apparatus consists of four scintillator planes arranged in two pairs. For the purpose of this analysis, we focus on a single scintillator plane as the only constraint on trajectory resolution between two detectors is position resolution in one detector. The plane consists of an array of $8\times8$ individual scintillator cells, totaling 64 cells per plane. Each cell measures 6.25~×~6.25~cm, resulting in a total active detection area of 50 × 50 × 2 cm for each plane.

Monoenergetic muons with an energy of 4 GeV (the average at sea level) were tested, and only small angles of incidence (< 0.25 rad) were considered to ensure that the muon traversed all four scintillators. Lastly, a conservative quantum efficiency of 10\% was used. 

Muons traversing these cells deposit energy, which is the primary observable for reconstructing their path. This analysis is focused on the position reconstruction accuracy within a single scintillation detector plane based on the energy deposition patterns from simulated muon events.

\subsection{Centroid Reconstruction Method}
\begin{figure}[htb!]
\centering
\includegraphics[width=0.9\columnwidth]{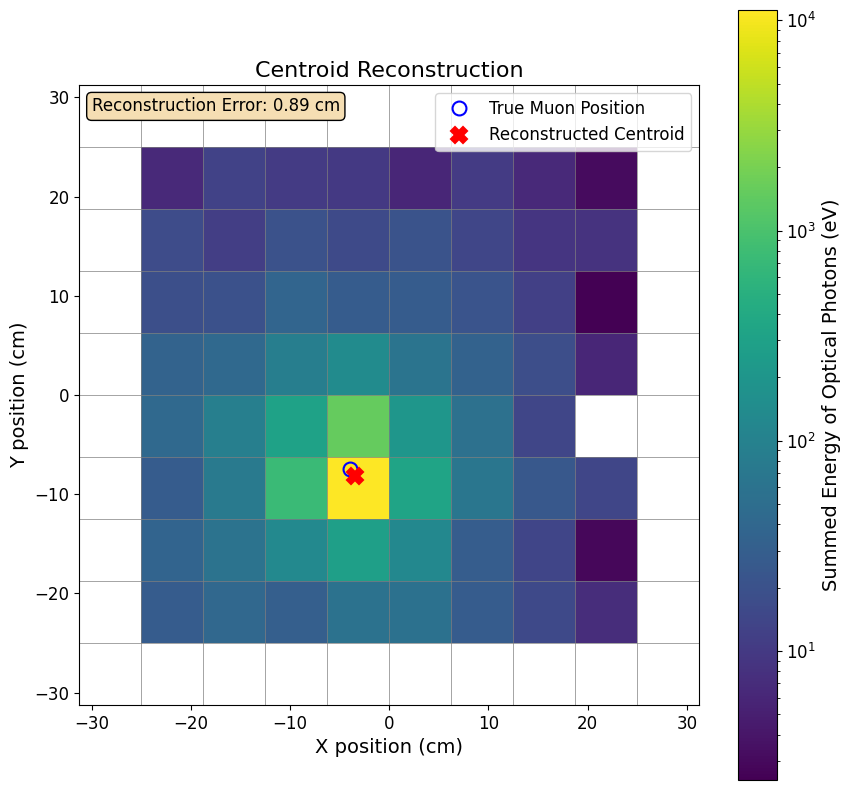}
\\[0.1em]
\includegraphics[width=0.9\columnwidth]{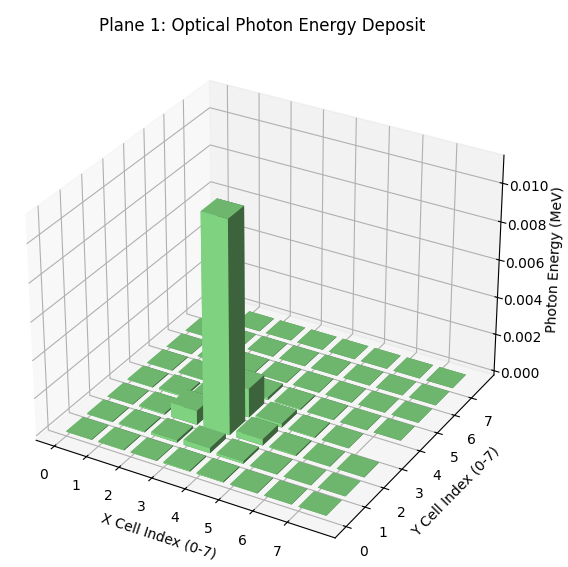}
\caption{Visualization of the centroid method for a single simulated muon event. (Top) The heatmap displays the summed energy of optical photons (eV) in each scintillator cell. (Bottom) Cell volume is proportional to the energy distribution in each cell.}
\label{fig:centroid_combined}
\end{figure}

The centroid method estimates the muon interaction position by calculating an energy-weighted average of the positions of the activated scintillator cells. The reconstructed position, $\vec{r}_{\text{recon}}$, is determined using Eq.~2 \cite{Landi:2002ke}:

\begin{equation}
\vec{r}_{\text{recon}} = \frac{\sum_i E_i \vec{r}_i}{\sum_i E_i}
\end{equation}

where $E_i$ represents the energy deposited in the $i$-th scintillator cell, and  $ \vec{r}_i$ represents the geometric center coordinates of that cell. This method assumes that cells closer to the true muon path will receive higher energy deposits and therefore weights their positions more heavily. Intuition for this assumption can be observed in Fig \ref{fig:centroid_combined}. The 2D and 3D reconstructions clearly convey the energy centroid being centered near the true position of the muon. 

\subsection{Maximum Likelihood Estimation}
The MLE method seeks to determine the muon interaction position $(x, y)$ that most likely explains the observed pattern of energy depositions in the scintillator cells \cite{10.1214/ss/1030037906, Richards_Jia}.

The detection of individual photons is a random, independent process, so the measured energy $M_k$ in each cell $k$ is assumed to be governed by Poisson statistics. The model, $\{E_k(x, y)\}$, then predicts the expected energy signal in cell $k$ for a given muon position. For a proposed muon position $(x, y)$, the model estimates geometric factors for path length energy distribution, optical attenuation, and a solid angle factor. 

In practice, maximizing the likelihood is numerically equivalent to minimizing the negative log-likelihood (NLL) function. For a Poisson distribution, the NLL is: 

\begin{equation}
\mathcal{L}(x, y) = - \sum_k \left( M_k \log(E'_k) - E'_k \right)
\end{equation}

where $M_k$ is the measured energy in cell $k$ and $E'_k$ is the model-predicted energy. 

The minimization of the NLL with respect to $(x, y)$ is performed numerically using the L-BFGS-B algorithm \cite{Liu1989}. This iterative optimization is initialized with a guess for the muon position derived from a centroid calculation and is constrained by the known physical boundaries of the detector plane. The $(x, y)$ coordinates that yield the minimum NLL are taken as the reconstructed muon position.

\subsection{Machine Learning Approach}
A feedforward neural network was developed to predict muon interaction coordinates $(x, y)$ based on engineered features derived from energy deposition patterns. The network employs a fully connected architecture designed to process spatial event characteristics.

The network processes a feature vector of 84 elements: 64 normalized energy depositions (1 per scintillator cell) and 20 engineered global features, including spatial moments up to fourth order, energy concentration metrics, directional asymmetries, and distribution characteristics. Prior to training, these global features were standardized using scikit-learn \cite{scikit-learn}.

The architecture consists of five fully connected layers with dimensions [512, 256, 128, 64, 32], chosen to progressively compress the feature space while maintaining sufficient capacity for spatial pattern recognition. Each layer is followed by ReLU activation. The final layer outputs the predicted $(x, y)$ coordinates directly.

The model was trained on 80\% of simulated events for 150 epochs using the AdamW optimizer \cite{kingma2017adam} with an initial learning rate of $10^{-3}$, selected for its improved generalization through decoupled weight decay. Mean squared error loss was employed to minimize the distance between predicted and true muon positions. Hyperparameters were optimized through validation set performance to minimize reconstruction error.

\section{Results and Analysis}
Each reconstruction method was evaluated on a test set of 5,000 simulated muon events using the energy deposition patterns recorded in the scintillator cells of the first interaction plane. This sample size balanced statistical significance with computational efficiency. The analysis indicates that all implemented methods achieve subpixel resolution.

\begin{table}[h]
\caption{Reconstruction Performance for 5,000 4 GeV Muons}
\centering
\resizebox{\columnwidth}{!}{%
\begin{tabular}{lccc}
\toprule
Method & Mean error (cm) & RMSE (cm) & Std dev (cm) \\
\midrule
Centroid       & 1.163 & 1.260 & 0.486 \\
MLE            & 0.678 & 0.777 & 0.380 \\
Neural network & 0.583 & 0.673 & 0.336 \\
\bottomrule
\end{tabular}%
}

\label{tab:comparison}
\end{table}

As shown in TABLE \ref{tab:comparison}, the accuracy of the reconstruction methods scales with their complexity. The centroid method achieved a mean error of 1.163 cm, equivalent to approximately 19\% of the scintillator cell width. The MLE method improved upon this with a mean error of 0.678 cm (11\% of cell width). The neural network produced the superior result, with a mean error of just 0.583 cm, which is less than 10\% of the cell width.

\begin{figure}[h!]
\centering
\includegraphics[width=0.90\columnwidth]{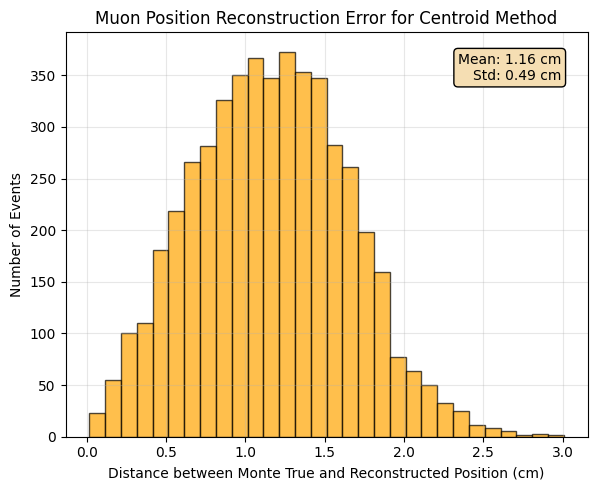}
\\[0.5em]
\includegraphics[width=0.90\columnwidth]{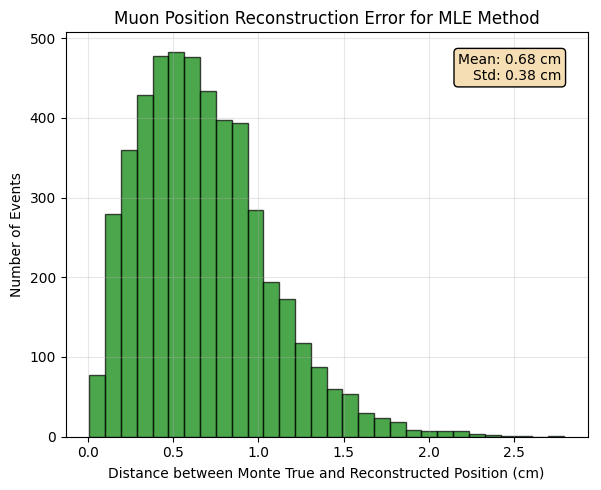}
\\[0.5em]
\includegraphics[width=0.90\columnwidth]{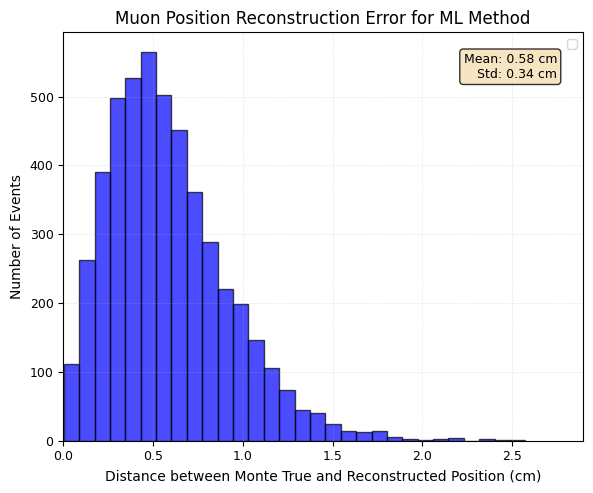}
\caption{Histograms of the position reconstruction error for the centroid (top), MLE (middle), and ML (bottom) methods. The $x$-axis represents the distance between the Monte Carlo true position and the reconstructed position in centimeters. The $y$-axis shows the number of events.}
\label{fig:histograms}
\end{figure}

These quantitative results are supported by the error distributions shown in Fig.~\ref{fig:histograms}. The histogram for the centroid method (top) is the broadest, corresponding to its largest standard deviation. By contrast, the distributions for the MLE (middle) and neural network (bottom) methods are progressively narrower and more sharply peaked, visually confirming their lower mean errors and smaller standard deviations.

The error distributions in Fig.~\ref{fig:histograms} are visibly skewed rather than Gaussian, with tails extending toward larger errors. This skewness likely stems from detector effects. Muons hitting near cell boundaries lose position information asymmetrically, and the discrete 6.25 cm cell size creates a natural bias. For practical muon tomography systems, this means reconstruction uncertainties cannot be treated as simple Gaussian errors. Instead, empirical error models based on the actual distributions should be used.

Given these results, implementing ML methods is a promising strategy for future scintillator-based detectors. Although the neural network is more computationally demanding than the other methods, the relatively low flux of cosmic muons at sea level (1 cm$^{-2}$ min$^{-1}$) provides sufficient processing time between events, making this more accurate method a practical choice.

\section{Conclusions}

This work systematically compared three position reconstruction algorithms (the centroid, MLE, and a neural network) for extending muon position reconstruction beyond pixel resolution. Our analysis shows that the use of ML via a neural network provides the best position resolution, achieving a mean error below 10\% of pixel width.

Improved position resolution is critical for reconstructing muon trajectories. Because the low flux of cosmic muons is a limiting factor in imaging speed, maximizing the precision of each measurement is essential. This precision leads to faster and more accurate imaging of high-Z materials. Improved speed and accuracy are vital for applications such as screening for shielded nuclear materials or verifying the contents of spent nuclear fuel casks.

Our analysis shows that with existing detector technology, algorithmic improvements alone can reduce position uncertainty to a level more than 10$\times$ smaller than the physical pixel size. Future work should validate these findings by simulating the broad energy and angular distributions of real-world cosmic muons. Furthermore, these results motivate the exploration of more advanced ML architectures to further improve reconstruction accuracy. 

In conclusion, we have demonstrated that achieving subpixel resolution on existing scintillator-based position detectors can be practically achieved and greatly reduced below the scale of existing pixel width. This methodology will lead to faster reconstruction of high-Z geometries with higher overall confidence. 

\section{Future Work}

A natural extension of the presented work is to test different cell geometries, 16 × 16 or 64 × 64 configurations for example, since more cells would provide more input data for the computational models, and therefore could lead to higher reconstruction accuracy. Additionally, these position reconstruction techniques should be extended to trajectory reconstruction in tomography detectors. Lastly, this methodology should be implemented in real-world detectors to test their effectiveness, especially because the presented simulation did not include a continuous muon energy spectrum or electronic noise.

\section{Acknowledgments}
This research was sponsored by the Laboratory Directed Research and Development Program of Oak Ridge National Laboratory, managed by UT-Battelle LLC for the US Department of Energy.

\bibliographystyle{ans}
\bibliography{bibliography}

@article{Landi:2002ke,
    author = "Landi, Gregorio",
    title = "{Properties of the center of gravity as an algorithm for position measurements}",
    eprint = "1908.04447",
    archivePrefix = "arXiv",
    primaryClass = "physics.ins-det",
    doi = "10.1016/S0168-9002(01)02071-X",
    journal = "Nucl. Instrum. Meth. A",
    volume = "485",
    pages = "698--719",
    year = "2002"
}

@article{10.1214/ss/1030037906,
author = {John  Aldrich},
title = {{R.A. Fisher and the making of maximum likelihood 1912-1922}},
volume = {12},
journal = {Statistical Science},
number = {3},
publisher = {Institute of Mathematical Statistics},
pages = {162 -- 176},
year = {1997},
doi = {10.1214/ss/1030037906},
URL = {https://doi.org/10.1214/ss/1030037906}
}

@inbook{Richards_Jia, edition={Third}, title={Chapter 8: Supervised Classification Techniques}, booktitle={Remote Sensing Digital Image Analysis}, publisher={Springer}, author={Richards, John A. and Jia, Xiuping}}

@article{PDG2020,
  author = {{Particle Data Group} and P. A. Zyla and others},
  title = {Review of Particle Physics},
  journal = {Progress of Theoretical and Experimental Physics},
  volume = {2020},
  number = {8},
  pages = {083C01},
  year = {2020},
  doi = {10.1093/ptep/ptaa104}
}

@article{scikit-learn,
  title={Scikit-learn: Machine Learning in {P}ython},
  author={Pedregosa, F. and Varoquaux, G. and Gramfort, A. and Michel, V.
          and Thirion, B. and Grisel, O. and Blondel, M. and Prettenhofer, P.
          and Weiss, R. and Dubourg, V. and Vanderplas, J. and Passos, A. and
          Cournapeau, D. and Brucher, M. and Perrot, M. and Duchesnay, E.},
  journal={Journal of Machine Learning Research},
  volume={12},
  pages={2825--2830},
  year={2011}
}

@misc{kingma2017adam,
      title={Adam: A Method for Stochastic Optimization}, 
      author={Diederik P. Kingma and Jimmy Ba},
      year={2017},
      eprint={1412.6980},
      archivePrefix={arXiv},
      primaryClass={cs.LG},
      url={https://arxiv.org/abs/1412.6980}, 
}

@article{doi:10.1098/rsta.2018.0052,
author = {Poulson, Dan  and Bacon, Jeff  and Durham, Matt  and Guardincerri, Elena  and Morris, C. L.  and Trellue, Holly R. },
title = {Application of muon tomography to fuel cask monitoring},
journal = {Philosophical Transactions of the Royal Society A: Mathematical, Physical and Engineering Sciences},
volume = {377},
number = {2137},
pages = {20180052},
year = {2019},
doi = {10.1098/rsta.2018.0052},

URL = {https://royalsocietypublishing.org/doi/abs/10.1098/rsta.2018.0052},
eprint = {https://royalsocietypublishing.org/doi/pdf/10.1098/rsta.2018.0052}
}

@Article{Bae2024,
author={Bae, JungHyun
and Montgomery, Rose
and Chatzidakis, Stylianos},
title={Momentum informed muon scattering tomography for monitoring spent nuclear fuels in dry storage cask},
journal={Scientific Reports},
year={2024},
month={Mar},
day={20},
volume={14},
number={1},
pages={6717},
issn={2045-2322},
doi={10.1038/s41598-024-57105-y},
url={https://doi.org/10.1038/s41598-024-57105-y}
}

@INPROCEEDINGS{8069919,
  author={Burns, J and Steer, C and Stapleton, M and Quillin, S and Boakes, J and Eldridge, C and Grove, C and Chapman, G and Lohstroh, A},
  booktitle={2016 IEEE Nuclear Science Symposium, Medical Imaging Conference and Room-Temperature Semiconductor Detector Workshop (NSS/MIC/RTSD)}, 
  title={Portable muon scattering tomography detectors for security imaging applications}, 
  year={2016},
  volume={},
  number={},
  pages={1-5},
  doi={10.1109/NSSMIC.2016.8069919}}

@article{BAE2024110240,
title = {Nuclear material accountancy using momentum-informed muon scattering tomography},
journal = {Annals of Nuclear Energy},
volume = {197},
pages = {110240},
year = {2024},
issn = {0306-4549},
doi = {https://doi.org/10.1016/j.anucene.2023.110240},
url = {https://www.sciencedirect.com/science/article/pii/S0306454923005595},
author = {JungHyun Bae and Rose Montgomery and Stylianos Chatzidakis}
}

@article{MOSES2011S236,
title = {Fundamental limits of spatial resolution in PET},
journal = {Nuclear Instruments and Methods in Physics Research Section A: Accelerators, Spectrometers, Detectors and Associated Equipment},
volume = {648},
pages = {S236-S240},
year = {2011},
issn = {0168-9002},
doi = {https://doi.org/10.1016/j.nima.2010.11.092},
url = {https://www.sciencedirect.com/science/article/pii/S0168900210026276},
author = {William W. Moses}
}

@Article{Liu1989,
author={Liu, Dong C.
and Nocedal, Jorge},
title={On the limited memory BFGS method for large scale optimization},
journal={Mathematical Programming},
year={1989},
month={Aug},
day={01},
volume={45},
number={1},
pages={503-528},
issn={1436-4646},
doi={10.1007/BF01589116},
url={https://doi.org/10.1007/BF01589116}
}
\end{document}